# What is the Shape Effect on the (Hyper)polarizabilities? A Comparison Study on the Möbius, Cyclic, and Linear Nitrogen-Substituted Polyacenes

Hong-Liang Xu, Zhi-Ru Li,* Fang-Fang Wang, Di Wu

*State Key Laboratory of Theoretical and Computational Chemistry, Institute of Theoretical Chemistry Jilin University, Changchun, 130023, China; E-mail: lzr@mail.jlu.edu.cn*

Kikuo Harigaya

*Nanotechnology Research Institute, AIST, Umezono 1-1-1, Japan*

Feng Long Gu *

*Department of Molecular and Material Sciences, Faculty of Engineering Sciences, Kyushu University, 6-1 Kasuga-Park, Fukuoka, 816-8580, Japan; E-mail: gu@cube.kyushu-u.ac.jp*

## Abstract

How does the framework shape influence the static polarizability ($\alpha_0$) and the first hyperpolarizability ($\beta_0$)? This work, for the first time, presents a comparison study at the MP2/6-31+G(d) level, by using the nitrogen-substituted polyacenes as models: the möbius strip with a knot, the cyclic strip without knot, and the linear strip. Opening the knot of the möbius strip to form the cyclic strip, it leads to the increase of the $\alpha_0$ from 268 au to 323 au and the $\beta_0$ value increases about three times from 393 (möbius) to 1049 au (cyclic). Further, opening the cyclic strip to form the linear strip, the $\alpha_0$ value increases from 323 au to 476 au. While the β0 value as well increases about three times from 1049 (cyclic) to 2814 au (linear). The changes in the static (hyper)polarizabilities are well explained by the geometrical differences among the möbius, cyclic, and linear nitrogen-substituted polyacenes.

## Introduction

The famous one-sided möbius strip has an interesting shape— closed strip with a knot. Some 100 years passed after Möbius' and Listing's mathematical explorations before Möbius topology caught the imagination of chemists.[1] Much effort is devoted to investigate the Möbius molecular structures[2-4] and these structures also exhibit special physical properties.[5-7] For example, the Hückel rules for aromaticity (4n+2 electrons) are no longer valid for Möbius annulenes and the Möbius ring with $4n$ $\pi$ electrons is aromatic. The unusual ring currents in Möbius annulenes are also particularly interesting.

The framework shape of Möbius strip molecule is very interesting. It is a closed strip with a knot exhibiting topological one-sided characteristic. How does the twisted framework of the Möbius strip molecule influence the nonlinear optical (NLO) property? This question is not solved yet, although quite many papers on the NLO properties of molecules and materials[8-20] have been published.

It is shown that doping is an important method to enhance the NLO responses by Champagne[15] and us.[21-29] In the present work, we use three nitrogen-substituted polyacenes with different framework shapes (a möbius strip with a knot and a cyclic strip without knot as well as a linear strip) as models to investigate the framework shape effect on the static polarizability and the first hyperpolarizability at the MP2/6-31+G(d) level.

**Computational Details**

The optimized geometric structures of möbius, cyclic, and linear strip nitrogen-substituted polyacenes with all real frequencies are obtained by using the B3LYP/6-31G(d) method. The static (hyper)polarizabilities are evaluated by a finite-field approach at the MP2/6-31+G(d) level. In our previous papers,[21-23] the MP2 results are close to those obtained from the more sophisticated correlation methods (for example, the QCISD[23]). The 6-31+G(d) basis set is sufficient for our purpose.[29] The magnitude of the applied electric field is chosen as 0.001 au[21-29] for the calculation of the (hyper)polarizabilities.

The dipole moment ($\mu_0$) and polarizability ($\alpha_0$) are defined as follows:

$$\mu_0 = (\mu_x^2 + \mu_y^2 + \mu_z^2)^{1/2} \tag{1}$$

$$\alpha_0 = \frac{1}{3}(\alpha_{xx} + \alpha_{yy} + \alpha_{zz}) \tag{2}$$

The static first hyperpolarizability is noted as:

$$\beta_0 = (\beta_x^2 + \beta_y^2 + \beta_z^2)^{1/2} \tag{3}$$

where $\beta_i = \frac{3}{5}(\beta_{iii} + \beta_{ijj} + \beta_{ikk}), i, j, k = x, y, z$.

All of the calculations were performed with the GAUSSIAN 03 program package.[30] The dimensional plots of molecular orbitals were generated with the GaussView program[31] (Gaussian, Inc. Pittsburgh, PA).

## Results and Discussions

### *A. Equilibrium Geometries*

Three molecules, the möbius strip with a knot, the cyclic strip without knot, and the linear strip, which are composed by seven nitrogen-substituted benzene rings. Their optimized geometric structures with all real frequencies are show in Figure 1.

From Table I and Figure 1, for möbius strip, there is the longest N-N (7-1) distance of 3.280 Å in distorted region as a consequence of the 180° twist. While in the nondistorted region the N-N distances (2.250 ~ 2.484 Å) are close to that of 2.268 Å for the cyclic strip. For the linear strip, except the very long N-N (7-1) distance of 13.280 Å between two ends by the effect of opening the ring, the other N-N distances are slightly longer by about 0.02 Å than that of the cyclic strip.

For möbius strip, the distribution range of C-N distance (1.221 ~ 1.460 Å) in distorted region (including unit 1, 2, 6 and 7) is larger than that (1.297 ~ 1.384 Å) in nondistorted region (including unit 3, 4 and 5). And these C-N distances (1.297 ~ 1.384 Å) in the nondistorted region of the möbius strip are close to those (1.349 ~ 1.350 Å) in the cyclic strip. For the linear strip, the distribution range of C-N distance (1.304 ~ 1.383 Å) in end region (including unit 1, 2, 6 and 7) is larger than that (1.334 ~ 1.349 Å) in non-end region (including unit 3, 4 and 5).

## B. The Static Polarizabilities and the First Hyperpolarizability

The electric properties of the möbius, cyclic and linear strip polyacenes calculated at the MP2 level are given in Table II. From Table II, the framework shape effect on the dipole moment ($\mu_0$) for the möbius and the cyclic as well as linear nitrogen-substituted polyacenes is obtained. The order of the $\mu_0$ values is 3.46 (möbius) < 5.33 (cyclic) ≈ 5.34 au (linear). It shows that opening the knot of the möbius strip the $\mu_0$ increases and opening the ring of the cyclic strip the $\mu_0$ is slightly changed. For the polarizabilities ($\alpha_0$), the order is 268.12 (möbius) < 323.38 (cyclic) < 476.33 au (linear), the framework shape obviously influences the polarizability.

Especially, the framework shape effect on the static first hyperpolarizability ($\beta_0$) of the möbius, cyclic, and linear nitrogen-substituted polyacenes is clearly shown. From Table II and Figure 2, the $\beta_0$ values are 393 (möbius), 1049 (cyclic) and 2814 au (linear). The framework shape effect on $\beta_0$ is seen, that is, opening the knot of the möbius strip to form the cyclic strip, the $\beta_0$ value increases about three times from 393 (möbius) to 1049 au (cyclic); further opening the cyclic strip to form the linear strip, the $\beta_0$ value again increases about three times from 1049 (cyclic) to 2814 au (linear).

In addition, the monotonic dependence of the $\beta_0$ value on electronic spatial extent ($R^2$) is also presented. The order of $R^2$ is 5686 (möbius) < 6023 (cyclic) < 20434 au (linear), which is consistent with that of $\beta_0$ (see Figure 2). The $\beta_0$ value increases with the $R^2$ values of the möbius, cyclic, and linear nitrogen-substituted polyacenes.

To further investigate the framework shape effect on $\beta_0$, we using the following two-level expression[32-34] to address the main influencing factors of $\beta_0$.

$$\beta_0 \propto \frac{\Delta\mu \cdot f_0}{\Delta E^3} \qquad (4)$$

In the above expression, $\beta_0$ is proportional to the difference of dipole moment between the ground state and the crucial excited state ($\Delta\mu$) and the oscillator strength ($f_0$), but inversely proportional to the third power of the transition energy ($\Delta E$).

The $f_0$, $\Delta\mu$ and $\Delta E$ are estimated by the CIS method with the 6-31+G (d) basis set and also listed in Table II for the möbius, cyclic, and linear strips. The $f_0$ values are 0.395 (möbius) < 0.624 (cyclic) < 2.031 (linear) and the $\Delta\mu$ values are 0.484 (möbius) < 1.039 (cyclic) >0.437 au (linear). The values of multipling $f_0$ by $\Delta\mu$ are more obvious and useful to see the effects of the different framework shapes. The radio of the $\Delta\mu \cdot f_0$ values is 1(0.191) for möbius: 3 (0.648) for cyclic: 5 (0.867 au) for linear strip polyacenes, which is close to the radio of $\beta_0$ to be 1(393) for möbius: 3 (1049) for cyclic: 7 (2814 au) for linear strip polyacenes. It is shown that the framework shape effect on the static first hyperpolarizability can be mainly understood by the changes of two important factors $f_0$ and $\Delta\mu$.

In the present work, we have obtained a valuable description of the framework shape effect on the static (hyper)polarizabilities for the möbius, cyclic, and linear nitrogen-substituted polyacenes. As a result, our investigation may evoke one's attention to the framework shape effect in designing new NLO compounds.

## Acknowledgment

This work was supported by the National Natural Science Foundation of China (No. 20573043, 20773046 and 20503010), and also by the Research and Development Applying Advanced Computational Science and Technology of the Japan Science and Technology Agency (ACT-JST).

**Table I.** The **C-N** and **N-N** Distances (Å) for the möbius, cyclic, and linear strip nitrogen-substituted polyacenes.

| C-N | Möbius | | cyclic | linear | |
|---|---|---|---|---|---|
| 1 | 1.290 | **distorted region** | 1.349 | 1.304 | **end part** |
|   | 1.445 |   |   | 1.383 |   |
| 2 | 1.305 |   | 1.349 | 1.322 |   |
|   | 1.403 |   |   | 1.359 |   |
| 3 | 1.297 | **nondistorted region** | 1.349 | 1.334 |   |
|   | 1.384 |   |   | 1.348 |   |
| 4 | 1.336 |   | 1.349 | 1.342 | **Non-end part** |
|   | 1.346 |   |   | 1.341 |   |
| 5 | 1.375 |   | 1.349 | 1.349 |   |
|   | 1.311 |   |   | 1.334 |   |
| 6 | 1.460 | **distorted region** | 1.349 | 1.360 | **end part** |
|   | 1.221 |   |   | 1.322 |   |
| 7 | 1.263 |   | 1.349 | 1.383 |   |
|   | 1.431 |   |   | 1.304 |   |
| **N-N** |   |   |   |   |   |
| 1-2 | 2.399 |   | 2.268 | 2.294 |   |
| 2-3 | 2.384 |   | 2.268 | 2.287 |   |
| 3-4 | 2.250 |   | 2.268 | 2.286 |   |
| 4-5 | 2.287 |   | 2.268 | 2.286 |   |
| 5-6 | 2.327 |   | 2.268 | 2.287 |   |
| 6-7 | 2.484 |   | 2.268 | 2.294 |   |
| 7-1 | 3.280 |   | 2.268 | 13.658 |   |

**Table II.** The values of the $\mu_0$, $\alpha_0$, $\beta_x$, $\beta_y$, $\beta_z$, $\beta_0$ and $R^2$ at MP2/6-31+g(d) level, $\Delta\mu$, $f_0$, $\Delta E$ at CIS/6-31+g(d) for the möbius, cyclic, and linear strip nitrogen-substituted polyacenes.

|  | $\mu_0$ | $\alpha_0$(au) | $\beta_x$(au) | $\beta_y$(au) | $\beta_z$(au) | $\beta_0$(au) | $R^2$(au) | $\Delta\mu$(au) | $f_0$ | $\Delta E$(eV) | $f_0 \cdot \Delta\mu$ |
|---|---|---|---|---|---|---|---|---|---|---|---|
| **Möbius** | 3.46 | 268.12 | -383 | 61 | 67 | 393 | 5686 | 0.484 | 0.395 | 6.144 | 0.191 |
| **Cyclic** | 5.33 | 323.38 | 0 | 1 | 1049 | 1049 | 6023 | 1.039 | 0.624 | 5.222 | 0.648 |
| **Linear** | 5.34 | 476.33 | 100 | -2812 | -1 | 2814 | 20434 | 0.437 | 2.031 | 5.634 | 0.867 |

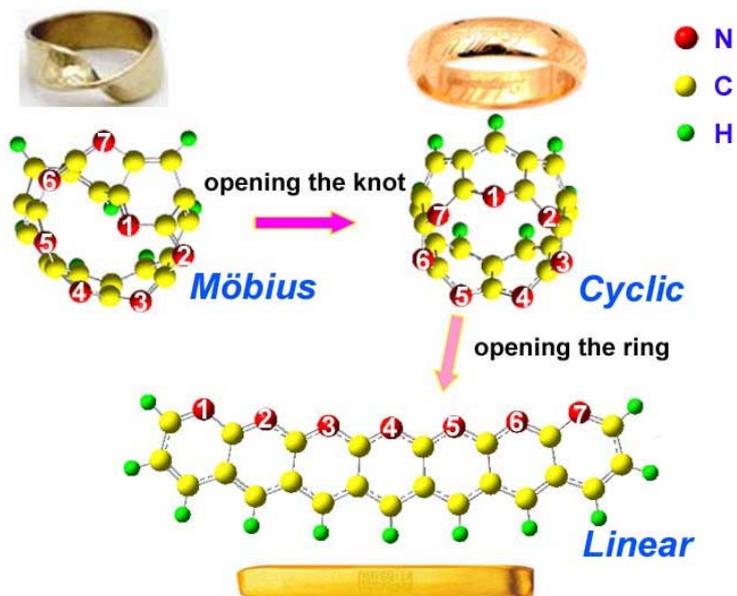

**Figure 1**. The geometric structures of the möbius, cyclic, and linear strip nitrogen-substituted polyacenes and their three analogues.

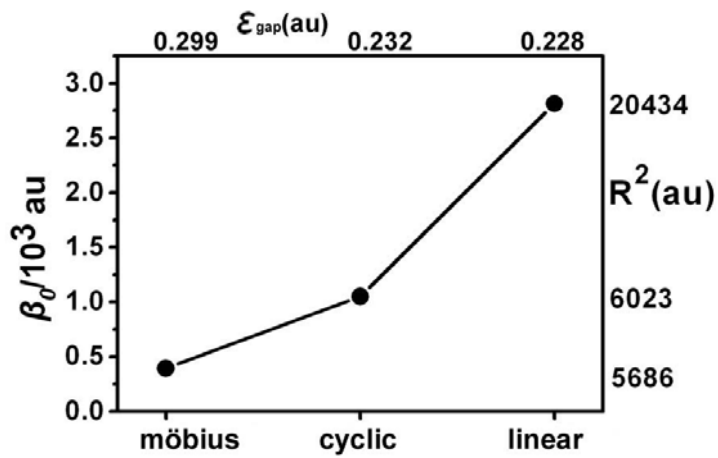

**Figure 2**. The $β_0$ values versus to the $R^2$ and $ε_{gap}$(LUMO-HOMO) for the möbius, cyclic, and linear strip nitrogen-substituted polyacenes.

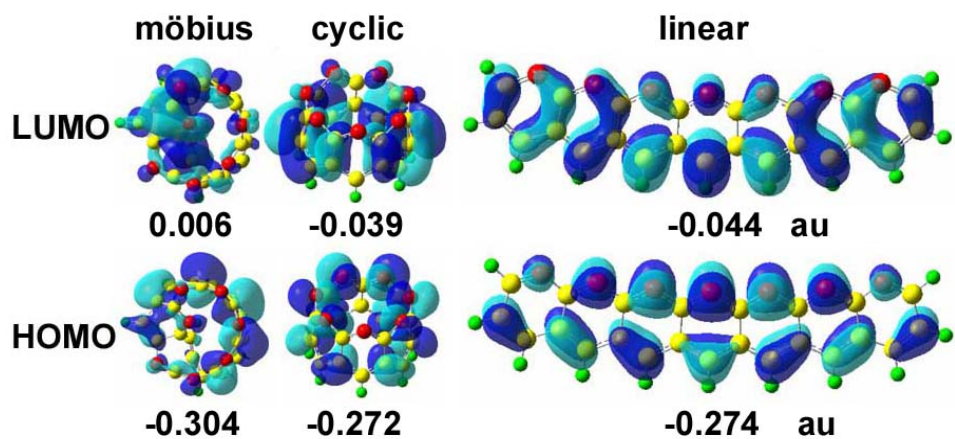

**Figure 3**. The **LUMO** and **HOMO** for the möbius, cyclic, and linear strip nitrogen-substituted polyacenes.